\documentclass[twocolumn,aps,pra,superscriptaddress,footinbib,floatfix]{revtex4-2}
\usepackage[utf8]{inputenc}

\usepackage{amsmath,amssymb,bm}
\usepackage{hyperref,xcolor,graphicx,siunitx}

\usepackage{tikzsymbols}
\usepackage{slashed}

\newcommand{\diag}{\mathrm{diag}}
\newcommand{\ket}[1]{|#1\rangle}

\newcommand{\braket}[2]{\langle #1|#2\rangle}
\newcommand{\braOket}[3]{\langle #1|#2|#3\rangle}

\begin{document}
\title{Chiral limit and origin of topological flat bands in twisted transition metal dichalcogenide homobilayers
}
\author{Valentin Cr\'epel}
\affiliation{Center for Computational Quantum Physics, Flatiron Institute, New York, New York 10010, USA}
\email{vcrepel@flatironinstitute.org}
\author{Nicolas Regnault}
\affiliation{Laboratoire de Physique de l'Ecole normale sup\'{e}rieure, ENS, Universit\'{e} PSL, CNRS, Sorbonne Universit\'{e}, Universit\'{e} Paris-Diderot, Sorbonne Paris Cit\'{e}, 75005 Paris, France}
\affiliation{Department of Physics, Columbia University, New York, NY 10027, USA}
\author{Raquel Queiroz}
\affiliation{Center for Computational Quantum Physics, Flatiron Institute, New York, New York 10010, USA}
\affiliation{Department of Physics, Columbia University, New York, NY 10027, USA}

\begin{abstract}
The observation of zero field fractional quantum Hall analogs in twisted transition metal dichalcogenides (TMDs) asks for a deeper understanding of what mechanisms lead to topological flat bands in two-dimensional heterostructures, and what makes TMDs an excellent platform for topologically ordered phases, surpassing twisted bilayer graphene. 
To this aim, we explore the chiral limits of massive Dirac theories applicable to $C_3$-symmetric moir\'e materials, and show their relevance for both bilayer graphene and TMD homobilayers. In the latter, the Berry curvature of valence bands leads to relativistic corrections of the moir\'e potential that promote band flattening, and permit a limit with exactly flat bands with nonzero Chern number. The relativistic corrections enter as a \emph{layer-orbit coupling}, analogous to spin-orbit coupling for relativistic Dirac fermions, which we show is non-negligible on the moir\'e scale. 
The Berry curvature of the TMD monolayers therefore plays an essential role in the flattening of moir\'e Chern bands in these heterostructures.
\end{abstract}

\maketitle


\section*{Introduction} 

Extremely narrow bands near magic angle in twisted bilayer graphene (TBG) are natural hosts for strongly correlated phenomena. 
At the core of the understanding of the TBG phase diagram nevertheless lie single-particle insights~\cite{bistritzer2011moire} that only weakly account for the full interactions of the system. 
This stems from a hierarchy of energy scales~\cite{bultinck2020ground} enabling to impose stronger symmetry constraints on theoretical models of TBG at only small costs in their experimental pertinence~\cite{song2021twisted}. 
The prime example of such physically relevant yet approximate model is the first chiral limit of TBG~\cite{tarnopolsky2019origin}, in which interlayer hopping is neglected where the graphene sheets stand furthest apart in the moir\'e unit cell. 
This extreme limit provides Landau-level like~\cite{wang2021chiral,estienne2023ideal} exact flat bands (EFBs)~\cite{ren2021wkb,watson2021existence,becker2023integrability} for a twist angle close to the experimental magic value~\cite{cao2018unconventional}. 
Relying on this structure, and adapting analytical results on ferromagnetism~\cite{sondhi1993skyrmions,girvin2000spin} and exact zero modes~\cite{haldane1983fractional,trugman1985exact,crepel2019matrix} from multi-component Hall systems, it is possible to rationalize the quantum anomalous Hall (QAH) state observed at filling $n=3$ in presence of aligned hBN~\cite{sharpe2019emergent,serlin2020intrinsic} and the fractional quantum Hall (FQH) states evidenced at low magnetic fields~\cite{xie2021fractional}.

Twisted transition metal dichalcogenide (TMD) homobilayers have also been identified as prominent platforms for strongly interacting topological phases~\cite{mak2022semiconductor} due to their predicted topological~\cite{wu2019topological} and extremely narrow~\cite{devakul2021magic,crepel2024bridging} topmost valence bands. 
Recently, Refs.~\cite{cai2023signatures,zeng2023integer} have reported clear signatures of strongly correlated physics using compressibility and optical measurements on twisted MoTe$_2$ homobilayers at twist angles $\theta \sim 3.5^\circ$. 
Independently, these two experiments have detected a robust spin-valley polarized QAH state at unit filling of the moir\'e unit cell~\cite{wu2019topological,devakul2021magic}, which hands down its ferromagnetism below unit filling to realize a half-metal~\cite{crepel2022anomalous} that acts as a precursor to the fractional Chern insulator (FCI), the zero magnetic field analog of the FQHE~\cite{neupert2011fractional,sheng2011fractional,regnault2011fractional,zhao2023fractional,crepel2024attractive}, measured at $2/3$ filling~\cite{cai2023signatures,zeng2023integer}.

At the moment, these moir\'e semiconductors lack the deep analytical structure offered by the approximate models available for TBG, preventing the same level of understanding. 
This absence of well controlled theoretical limit for TMDs does not stem from a fundamental opposition with TBG. 
Indeed, the only difference between moir\'e TMDs and TBG is a large $C_2$-symmetry breaking mass term gapping out the Dirac cones, analogous to, but much stronger than, the effect of aligned hBN on TBG~\cite{zhang2019twisted}.

In this article, we comprehensively explore \emph{all} the chiral limits of perturbed Dirac field theories applicable to $C_3$-symmetric moir\'e materials. 
We find that that only two exist, one of which is guaranteed to feature exact flat bands (EFBs)~\cite{atiyah1963index,parhizkar2023generic,tarnopolsky2019origin}. 
For massless fermions, they reproduce the two chiral limits identified in TBG~\cite{tarnopolsky2019origin,song2021twisted}. 
Extending these limits to massive Dirac theories, we observe that the one possessing topological EFBs perfectly captures the physics of twisted TMD homobilayers. 
In fact, we argue that, due to larger corrugation effects, TMDs are a better realization of the first chiral limit than TBG itself.

This first massive chiral limit explains the emergence of topological flat bands in TMD homobilayers. Moreover, our approach qualitatively captures the special angles $\theta \sim 3.5^\circ$ where experiments have observed zero-field FCIs~\cite{cai2023signatures,zeng2023integer}. 
The topological character of the flat bands can be understood as a consequence of layer-orbit coupling, the analog of spin-orbit coupling in the standard relativistic Dirac theory.
Contrary to the relativistic case however, we show that this term, which to our knowledge does not appear elsewhere, is non-negligible on the moir\'e scale since the Dirac velocity divided by the moir\'e period $v/a_m \sim 10$meV is of similar magnitude as the interlayer hybridization scale.

Index theorems~\cite{atiyah1963index,parhizkar2023generic,crepel2024topologically} guarantee that the flatness of the bands in the chiral limit are, to first order, immune to both gauge and potential disorder akin to the zeroth Landau level of graphene~\cite{kailasvuori2009pedestrian}. The existence of this limit, even when fine-tuned, can therefore explain how delicate correlated phases such as FCIs appear in twisted TMDs near their first magic angle despite their disorder and strain.

\section*{Results and Dicussion}

\subsection*{Exact flat bands and chiral anomalies}

To guide our search for EFBs and magic angles, we build on the physics of TBG, wherein bands closest to charge neutrality perturbatively flatten when the velocity of the Dirac cones vanishes~\cite{bistritzer2011moire}, and become fully degenerate under slight modifications of the model that provide a chiral anomaly with non-trivial index~\cite{parhizkar2023generic}. 
Here, we briefly review these arguments and characterize the most general chiral symmetry applicable to $C_3$ symmetric moir\'e materials with Dirac cones.

We consider a two-dimensional heterostructure composed of two layers of massive or massless Dirac materials that hybridize with one another via slowly varying coupling terms on a large moir\'e scale $a_m$ set by the interlayer lattice mismatch and twist, as described by the generic Hamiltonian 
\begin{equation} \label{eq_originalmodel} 
h (r)  = \begin{bmatrix} \sigma^\mu D_\mu^+  & T^\dagger(r) \\ T(r) & \sigma^\mu D_\mu^-  \end{bmatrix} , 
\end{equation}
where $D_\mu^\pm = [\pm \delta \mu, i(v \pm \delta v) \partial_1, i (v \pm \delta v) \partial_2 , (m \pm \delta m)]$ includes different velocities $v\pm\delta v$, masses $m\pm \delta m$ and charge neutrality points $\pm \delta\mu$ for the two layers; while the slowly-varying hybridization matrix is for the moment left unconstrained and parameterized by $T(r) = (t_j + i \lambda_j) \sigma^j - (t_0 + i\lambda_0) \sigma^0$ with $j=1,2,3$. 
The $\sigma^\mu$ Pauli matrices represent the spinor structure of the low-energy Dirac fermions, which are distinguished by a layer pseudo-spin $\tau^\mu$ providing the additional block structure of Eq.~\ref{eq_originalmodel} with $\tau^3$ defining the layers.
Using the gamma matrices $\gamma^0 = \tau^1 \sigma^0$, $\gamma^{j=1,2,3} = i\tau^2 \sigma^j$, and $\gamma^5 = i \gamma^0 \gamma^1 \gamma^2 \gamma^3$, the action corresponding to $h$ can be compactly recast as (see Supplementary Note 1)
\begin{subequations} \label{eq_fermionic2plus1action}
\begin{equation} \label{eq:action}
\mathcal{S} = \int {\rm d}^3 x \, \bar\psi \left[ i\slashed{D} + M \right] \psi ,
\end{equation}
where $x^0 = vt$ represents time, $\psi$ is the fermionic operators with conjugate $\bar\psi = \psi^\dagger \gamma^0$, and we have defined
\begin{align}
\slashed{D} &= \gamma^a [v \partial_a + i t_a \gamma^5 + i \lambda_a (i\gamma^3) ] - \delta v  \gamma^3 \gamma^5 \gamma^j \partial_j  , \label{eq_slashedD} \\
M &= \delta m + \delta \mu \gamma^0 \gamma^3 \gamma^5 - m \gamma^3 \gamma^5 - t_3 \gamma^3 - i\lambda_3 \gamma^5 ,
\end{align} \end{subequations}
with $a=0,1,2$ and $j=1,2$. 
Our search for EFBs in this generic model relies on a necessary and a self-consistent condition that we now detail.

First, when the Dirac cones' position is locked by crystalline symmetries, such as $C_3$ for TBG, perturbative flattening of the bands due to a vanishing Dirac cone velocity serves as a pre-requisite to EFBs. 
Such vanishing generically occurs upon tuning a single microscopic parameter, \textit{e.g.} the twist angle, when the co-dimension of the velocity operator
is equal to one in the space of possible parameters~\cite{sheffer2023symmetries}. 
The co-dimension of the velocity depends on the symmetries of the Hamiltonian, and in particular cannot equal one in absence of particle-hole (PH) symmetry~\cite{sheffer2023symmetries}.
Our necessary condition for finding EFBs is therefore that the model Eq.~\ref{eq_fermionic2plus1action} possess a PH symmetry.

Turning to the self-consistency condition, let us assume the existence of EFBs in our problem. 
Projected onto these flat bands, the action of Eq.~\ref{eq_fermionic2plus1action} loses its time dependence and reduces from $(2+1)$ to $(2+0)$ dimensions, \textit{e.g.} via $\bar \psi (v \gamma^0 \partial_0) \psi = 0$ in the case of degenerate flat bands at zero energy~\cite{parhizkar2023generic}. 
The previously imposed PH symmetry provides an operator exchanging positive and negative energy states that anti-commutes with the Hamiltonian and plays the role of a chiral symmetry. 
Since chiral Dirac fermions in even dimensions exhibit a chiral anomaly, the model in Eq.~\ref{eq_originalmodel} in presence of EFBs must satisfy additional constraints.
More precisely, the chiral anomaly forces the effective gauge fields descending from $t_a$ and $\lambda_a$, respectively carried by $\gamma^5$ and $\gamma^3$ in Eq.~\ref{eq_fermionic2plus1action}, to yield a non-zero integer Atiyah-Singer index~\cite{atiyah1963index,fujikawa1979path}. 
This is the self-consistency condition for EFBs, first pioneered in Refs.~\cite{parhizkar2023generic}, that we will use. 
This self-consistency condition can be intuitively understood as a Landau-like quantization, since it restrains the flux of chiral gauge fields on elementary real-space patches to integer values; suggesting a deeper connection between such flat bands and generic Landau levels~\cite{estienne2023ideal}.

In fact, the non-trivial Atiyah-Singer index of graphene's Landau levels is the reason why they remain exactly flat even under a non-uniform magnetic field or weak scalar potentials~\cite{kailasvuori2009pedestrian,kawarabayashi2009quantum}.
The robustness against varying gauge fields and small perturbations provided by the index theorem is the hidden underlying reason explaining why the phase diagram of TBG is so remarkably reproducible, and why it is so closely connected to the first chiral limit in spite of non-negligible AB-hoppings. Deriving a chiral limit for moir\'e semiconductors does not only provide a fine tuned EFB model for these materials, but rather argues that such EFBs are resilient to local perturbations through the same anomaly-protection present in TBG.

Relying on index theorems~\cite{atiyah1963index}, PH/chiral symmetries provide self-consistent conditions enabling to localize EFBs in models of coupled Dirac cones~\cite{parhizkar2023generic}. 
We now exhaustively use this method on our original model Eq.~\ref{eq_fermionic2plus1action} assuming a moir\'e pattern with $C_3$ symmetry.

\subsection*{The chiral limits of $C_3$ symmetric moir\'e} 

We have enumerated all possible PH/chiral symmetries compatible with $v\neq 0$ or $\delta v \neq 0$ that our model may possess \emph{after dimensional reduction}, see Supplementary Note 2 that includes Refs.~\cite{bernevig2021twisted,crepel2023chiral,song2021twisted,bistritzer2011moire,tarnopolsky2019origin}. 
Focusing on the most relevant case of $C_3$-symmetric moir\'e homobilayers and assuming the lowest harmonics of $T(r)$ are non-zero~\footnote{Higher harmonics respecting the $C_3$-symmetry can be included to the hybridization $T(r)$. They will not change the possible chiral symmetries of the model, but can alter the value of the magic $\alpha_c$~\cite{becker2023integrability}.}, we have demonstrated that only two inequivalent PH/chiral symmetries can emerge in our setup. 
They correspond to the two inequivalent chiral limits previously identified in TBG~\cite{tarnopolsky2019origin,song2021twisted}.

\begin{figure}
\centering
\includegraphics[width=\columnwidth]{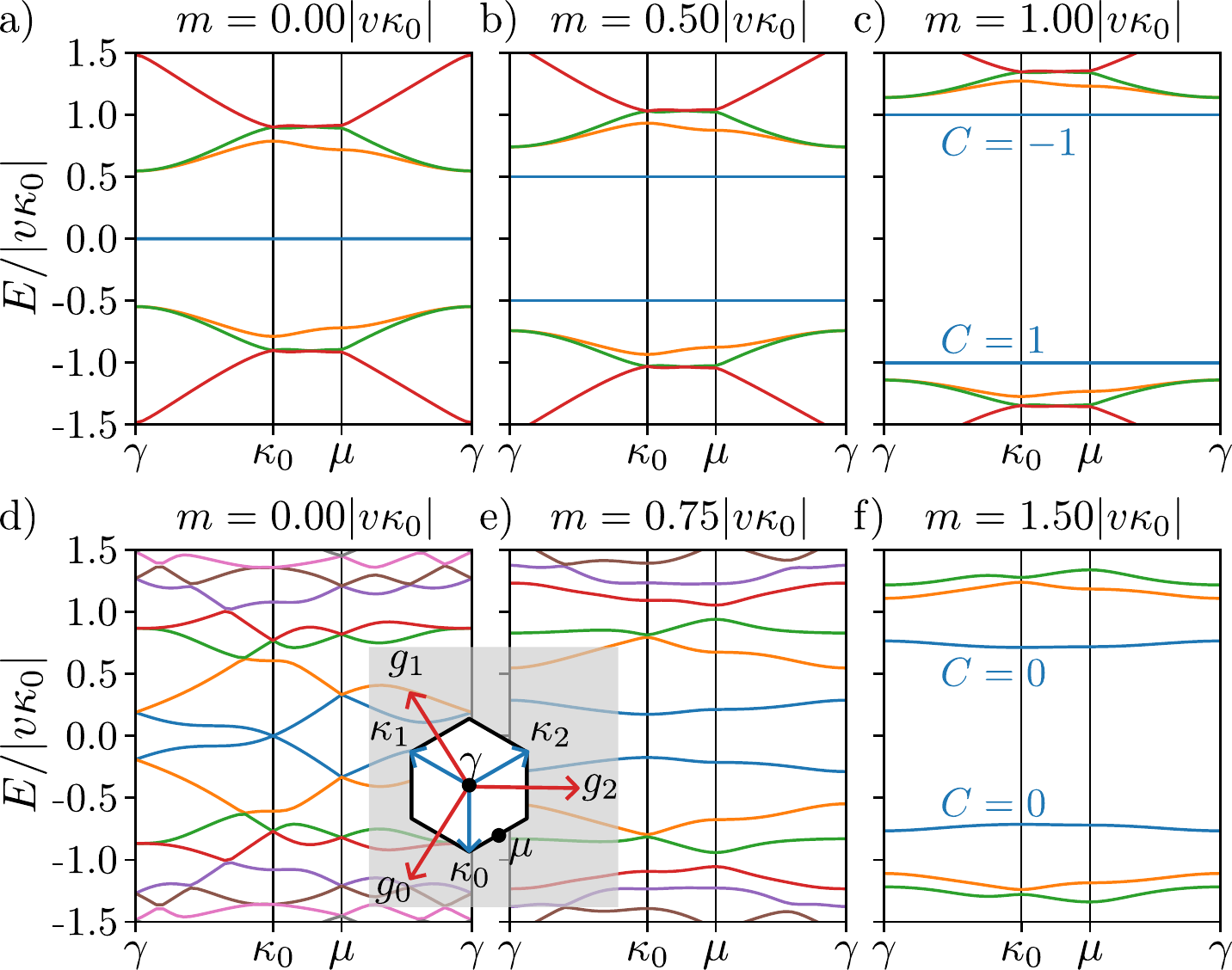}
\caption{
Band structure in the two sole massive chiral limits of $C_3$ symmetric moir\'e material with Dirac cones in presence of an additional mass $m$. 
The first (a-c) and second (d-f) chiral limits reproduce those of TBG when $m=0$ (a, d) and feature isolated narrow bands for large values of the mass (c, f). 
Exact flat bands with opposite Chern numbers appear in the first chiral limit (a-c) for any $m$ due to a chiral anomaly with non-zero Atiyah-Singer index -- analogous to the zero-th Landau level of graphene in presence of sublattice potential difference. 
We used $\alpha = \alpha_c$ in (a-c, see Eq.~\ref{eq_firstchiralmodel}), $\alpha = 1.2\alpha_c$ and $r=0$ in (d-f, see Eq.~\ref{eq_secondchiralmodel}), and the inset shows high symmetry points in reciprocal space, with $g_n = \kappa_n - \kappa_{n-1}$.
}
\label{fig_massivechiralmodel}
\end{figure}

The first PH/chiral symmetry, $\gamma^0 \gamma^3 \gamma^5$, becomes a faithful anti-commuting symmetry of the dimensionally reduced $(2+0)$d action provided that it has no mass terms $M=0$ and that its hybridization assumes the form 
\begin{equation} \label{eq_firstchiralmodel}
T^{\rm st}(r) =  t \sum_{n=0}^2 e^{i (\kappa_n \cdot r)} \begin{bmatrix} 0 & \omega^n \\ \omega^{-n} & 0 \end{bmatrix} , \quad \alpha = \frac{t}{|v \kappa_0|} ,
\end{equation}
with $\omega = \exp(2 i \pi/3)$ and $\kappa_n$ the clockwise $2n\pi/3$ rotation of the moir\'e Brillouin zone corner $\kappa_0$ along $(-\hat y)$, see inset of Fig.~\ref{fig_massivechiralmodel}. 
This $(2+0)$d model exactly reproduces the first chiral limit of TBG~\cite{tarnopolsky2019origin}, whose spectrum only depends on the dimensionless parameter $\alpha$ that is inversely proportional to the twist angle $\theta$. It is known to feature exact flat bands for equally spaced values of $\alpha$~\cite{ren2021wkb,watson2021existence,becker2023integrability}. 
This result was recovered in Ref.~\cite{parhizkar2023generic} by direct evaluation of the Atiyah-Singer index corresponding to Eq.~\ref{eq_firstchiralmodel}, providing integer values $n$ for $\alpha = n\alpha_c$ (see also Supplementary Note 3 that includes Refs.~\cite{parhizkar2023generic,fujikawa1979path,tarnopolsky2019origin}). 
The largest value $\alpha = \alpha_c = 1/\sqrt{3}$ gives an estimate of the first magic angle $\theta_c = 3 \sqrt{3} t a/(4 \pi v) \simeq 1.1^\circ$~\cite{parhizkar2023generic} close to the experimentally observed value~\cite{cao2018unconventional}.

The second PH/chiral symmetry available, $\gamma^0$, requires the $(2+0)$d theory to have $\delta \mu = \delta m = m=0$ and a specific hybridization matrix
\begin{equation} \label{eq_secondchiralmodel}
T^{\rm nd}(r) =  t \sum_{n=0}^2 e^{i (\kappa_n \cdot r)} \begin{bmatrix} 1+\epsilon & 0 \\ 0 & 1-\epsilon \end{bmatrix} , 
\end{equation}
with $\epsilon$ a real parameter. 
When the system possess an additional $C_2$ symmetry exchanging the layers' sublattices, $\epsilon=0$ and the previous form exactly matches the second chiral limit of TBG, describing a perfect metal in which all bands are connected~\cite{bernevig2021twisted,song2021twisted}. 
We show in the Supplementary Note 3 that the Atiyah-Singer index vanishes in this second chiral limit, preventing the appearance of exact flat bands.

Using generic principles and starting from the generic coupled Dirac Hamiltonian Eq.~\ref{eq_originalmodel}, we have identified the two sole chiral limits of $C_3$-symmetric moir\'e materials, only one of which provides exact flat bands due to its non-zero Atiyah-Singer index. 
The obtained chiral limits reproduce those identified in TBG, which bolsters our approach but also highlights that all EFBs of $C_3$-symmetric moir\'e materials feature TBG-like behaviors, in the sense that they derive from the same chiral anomaly.

\subsection*{Consequences for moir\'e semiconductors}

Moir\'e semiconductors can be described using the $(2+1)$d action in Eq.~\ref{eq_fermionic2plus1action} with $m \neq 0$. 
At first sight, this seems to preclude the realization of the chiral limits identified above (Eqs.~\ref{eq_firstchiralmodel} and~\ref{eq_secondchiralmodel}) since they both require $m=0$ in the dimensionally reduced action in $(2+0)$d where the chiral anomaly occurs. 
Fortunately, these two apparently conflicting theories do not only differ through their mass, but also through their dimensionality. 
We can thus devise a more general dimensional reduction scheme, compatible with particle-hole symmetry, that transforms the semiconducting $(2+1)$d action with non-zero mass to a massless $(2+0)$d action. 
This is achieved by projecting the $(2+1)$d action onto flat bands at $\pm m$ using $\bar \psi v \gamma^0 \partial_0 \psi =  m \psi^\dagger (\gamma^0 \gamma^3 \gamma^5) \psi$, which exactly compensates for the mass terms and yields massless effective $(2+0)$d theories and enables to realize the chiral limits obtained above.

Transposing our previous discussion, we predict that the first massive chiral limit, featuring the hybridization matrix Eq.~\ref{eq_firstchiralmodel} and a mass $m$, still has exact flat bands located at energies $\pm m$. 
We numerically checked this fact for different $m$, as shown in Fig.~\ref{fig_massivechiralmodel}a, where we clearly see the originally degenerate flat bands with non-trivial and opposite Chern number splitting to $\pm m$ energies. 
We have also checked that, as in chiral TBG~\cite{wang2021chiral}, these flat bands are idea (see Supplementary Note 4 that includes Refs.~\cite{wang2021chiral,estienne2023ideal,tarnopolsky2019origin,crepel2018matrix}), \textit{i.e.} equivalent to the physics of a Landau level with non-uniform magnetic field~\cite{estienne2023ideal}.
For small $m$, this describes the effect of an aligned hBN substrate on chiral TBG~\cite{zhang2019twisted}. 
The Chern numbers of the flat bands are inherited from those obtained for TBG in the first chiral limit, and can be understood by analogy with the zero-th Landau levels of graphene, which also split into two groups that can be fully localized on opposite sublattices. 
Upon adding a sublattice potential difference ---  equivalent to the mass term in Eq.~\ref{eq_originalmodel} --- these groups with pair-wise opposite Chern number acquire finite and opposite energies. 
While the second chiral limit cannot realize EFBs, we nevertheless highlight that the additional mass term considered here gaps out the perfectly connected bands obtained for $m=0$, and, for sufficiently large values, produces narrow and isolated trivial bands around charge neutrality (see Fig.~\ref{fig_massivechiralmodel}b).

Compared to TBG, both the first and the second chiral limit yields isolated narrow bands enhancing the interaction effects providing a fertile playground for the emergence of strongly correlated phases. The main difference lies in the topology of these narrow bands: they carry a non-zero Chern number in the first chiral limit and are topologically trivial in the second chiral limit. We finally notice that our method does not work for heterobilayers, for which all four coefficients ($v$, $\delta v$, $m$, $\delta m$) are non-zero in the (2+1)d theory and cannot be simultaneously accounted for in the dimensional reduction (see Supplementary Note 2), which suggests that no EFBs exists for semiconducting heterobilayers.

\subsection*{Dirac to Schr\"odinger, and layer-orbit coupling}

In presence of large gaps between the conduction and valence bands, an alternative description of the physics of $p$/$n$-doped moir\'e semiconductors should be possible in terms of hole/electron degrees of freedom only. We now derive such description by downfolding our Dirac theory using second order perturbation in the large gap $m$. We focus on the first massive chiral limit, for it features flat bands and non-trivial topology, and obtain an effective hole-like continuum model describing bands below the band gap. 
We find a crucial term, neglected in previous works, that differentiates between gapped Dirac physics and simple quadratic bands and provides a simple explanation for the non-trivial topology of the flat bands.

Up to the moir\'e potential terms, the derivation is identical to the transformation of the massive Dirac equation into the Schr\"odinger equation with quadratic dispersion. 
In relativistic settings, small corrections that appear, such as spin-orbit coupling, are inversely proportional to the speed of light. Usually, they can be safely discarded. 
For moir\'e materials, they cannot. These corrections are inversely proportional to the Dirac velocity, but $\alpha = \frac{t}{|v\kappa_0|} \sim 1$ implies that they are of the same order as the typical moir\'e hybridization strength.

Including deviations from the first chiral limit using a hybridization function $T = T^{\rm st} + \frac{|v\kappa_0|}{2m} \beta T^{\rm nd}$, where $\beta$ plays a role analogous to $w_0/w_1$ in the usual language of TBG~\cite{bistritzer2011moire,bernevig2021twisted}, standard second order perturbation theory (see Supplementary Note 5 that includes Refs.~\cite{wu2019topological,crepel2023topological}) gives the effective continuum theory for the valence bands of the moir\'e semiconductor
\begin{subequations} \label{eq_downfoldedmodelwithgradient} \begin{align}
\Tilde{h} (r) & = \frac{\hbar^2}{2 m^*} \nabla^2 \tau^0 - 2 V \sum_{n=0,1,2} \cos (g_n \cdot r + \tau^3 \psi) \\ 
& - i W \sum_{n=0,1,2} e^{i \kappa_n\cdot r} \left[ 1 + i \beta + \frac{2 \lambda (\kappa_n \times \nabla)}{|\kappa_0|^2} \right] \tau^- + hc , \notag 
\end{align}
with $g_n = \kappa_{n} - \kappa_{n-1}$ (see inset of Fig.~\ref{fig_massivechiralmodel}), and the global $i$ factor in front of the interlayer-hopping can be gauged away. The coefficients of the downfolded model are inferred from those of the first chiral limit
\begin{equation}
\frac{\hbar^2}{2 m^*} = \frac{v^2}{2m} , \; W = \frac{\alpha |\hbar \kappa_0|^2}{2m^*}  , \; V = \alpha W ,  \; \psi = \frac{2\pi}{3} , \; \lambda = 1 . 
\end{equation} \end{subequations}
Thus, up to an overall scaling coefficient, this Hamiltonian only depends on two dimensionless constants $(\alpha,\beta)$, or only one if we focus on the first chiral limit where $\beta = 0$. 
The parameters in Eq.~\ref{eq_downfoldedmodelwithgradient} can be combined in magic rule for moir\'e bilayers with gapped Dirac cones 
\begin{equation} \label{eq_ruleformagic}
\frac{\hbar^2 |\kappa_0|^2}{2 m^*} \frac{V}{W^2} = 1  \quad \text{or} \quad E_{\rm kin} E_{\rm intra} = E_{\rm inter}^2 ,
\end{equation}
where $E_{\rm kin / inter/intra}$ stand for the typical kinetic, inter- and intra-layer potential energies on the moir\'e scale. 
The rule Eq.~\ref{eq_ruleformagic} determining the magic angle is exact for the massive chiral model with $\beta =0$.
Beyond this model, we speculate that the second formulation in Eq.~\ref{eq_ruleformagic} still provides a good rule of thumb to estimate the twist angle at which correlated physics appears in moir\'e semiconductors (see Supplementary Note 6 that includes Refs.~\cite{devakul2021magic,morales2023pressure}).

\begin{figure}
\centering
\includegraphics[width=\columnwidth]{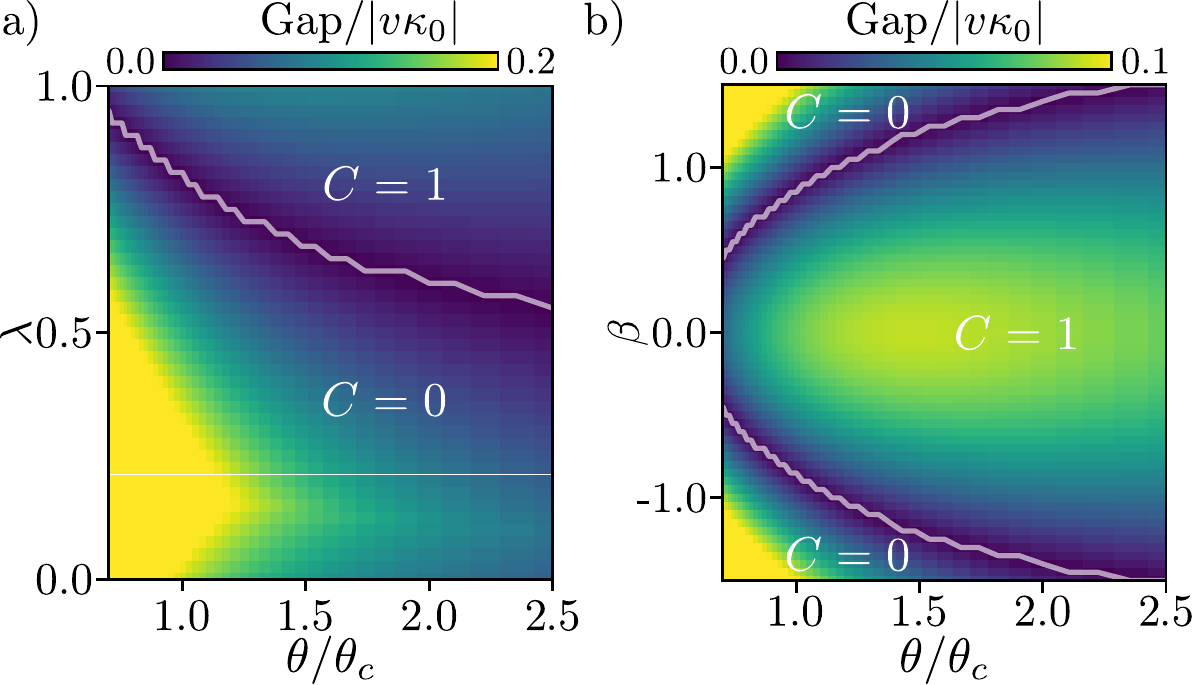}
\caption{
Topological phase diagram of chiral moir\'e semiconductors. 
(a) Upon turning off the layer-orbit coupling $\lambda$ --- the analog of spin-orbit coupling in the relativistic Dirac theory, which is \textit{non-negligible} for moir\'e materials --- the topmost valence band from the first chiral limit (Fig.~\ref{fig_massivechiralmodel}a) loses its topological character. All parameters except $\lambda$ are fixed by Eq.~\ref{eq_downfoldedmodelwithgradient} and $\beta = 0$. 
(b) Deviations away from the chiral limit (Eq.~\ref{eq_downfoldedmodelwithgradient} with $\lambda = 1$), measured by $\beta$, do not spoil the physics of the first massive chiral limit for a wide range of parameters. In both panels, $m=2|v\kappa_0|$ was used, and $\theta / \theta_c = \alpha_c / \alpha$ since $\alpha$ is inversely proportional to the twist angle.
}
\label{fig_ScanEpsilon}
\end{figure}

The main difference between Eq.~\ref{eq_downfoldedmodelwithgradient} and previous continuum models for moir\'e semiconductors~\cite{wu2019topological,pan2020band,devakul2021magic} is the advertised layer-orbit coupling $\lambda$. This is a non-negligible analog of spin-orbit coupling in relativistic systems.
We now stress that this layer-orbit coupling, necessary for a complete downfolding of the Dirac theory, is crucial for the emergence of non-trivial topological properties in moir\'e semiconductors. 
For this purpose, we consider Eq.~\ref{eq_downfoldedmodelwithgradient} in the first chiral limit ($\beta = 0$) and consider the effects of turning off $\lambda$. 
In Fig.~\ref{fig_ScanEpsilon}a, we plot the gap between the two topmost valence bands as a function of twist angle, measured by $\alpha_c/\alpha = \theta/\theta_c$, and $\lambda$. 
We observe a topological gap closure as $\lambda$ goes from its natural value, one, to zero. 
This proves the crucial role of layer-orbit coupling in Eq.~\ref{eq_downfoldedmodelwithgradient} to capture the topological character of the exact flat bands obtained in the first massive chiral limit (Fig.~\ref{fig_massivechiralmodel}a).

Including deviations from the first chiral model using $\beta \neq 0$, we can also drive a topological gap closure for sufficiently large values of $\beta$ (Fig.~\ref{fig_ScanEpsilon}b). 
The topological character of the topmost valence band obtained in the first chiral limit remains unchanged in a large regime of parameters. 
For instance, the transition to topologically trivial bands occurs for $\beta \simeq 0.8$ at the magic angle $\alpha=\alpha_c$.

\subsection*{Chirality of TMD homobilayers} 

We now relate our theory to the physics of twisted $K$-pocket semiconducting TMD homobilayers, the prime example of exfoliable gapped Dirac cone materials. We argue that, due to corrugation effects, these bilayers lie close to the magic limit than TBG.

In their monolayer form, the considered TMDs display a direct gap between electron and hole pockets at the corner of the Brillouin zone, which can be described as massive Dirac cones of gaps $m \sim 1-\SI{2}{\electronvolt}$ with conduction/valence orbitals mostly of the $d_{z^2}$/$d_{\pm} = d_{x^2-y^2} \pm i d_{xy}$ type near the $\pm K$ corners of the monolayer Brillouin zone. 
The minimal $k \cdot  p$ model describing the twisted TMD homobilayers is thus precisely of the form of Eq.~\ref{eq_originalmodel}. 
The symmetry of these orbitals determine the form of the interlayer hybridization at high-symmetry stacking points. 
We focus on R-type (or AA) stacking of the bilayers whose interlayer tunneling matrix $T$, determined by the symmetry of the bilayer in the lowest-harmonics approximation, is precisely in the form considered above $T = T^{\rm st} + W \beta T^{\rm nd}$,~\cite{tong2017topological}, with $\beta$ the ratio between direct interlayer hopping between $d_{\pm}$ orbitals and $W$.

Whether the first massive chiral model correctly captures the universal physics of these semiconducting TMD homobilayers only depends on the magnitude of $\beta$.
We now estimate the ratio $\beta$ using density functional theory results obtained in the literature for specific TMDs (WSe$_2$ and MoTe$_2$)~\cite{devakul2021magic,reddy2023fractional,wang2024fractional}. 
In these references, large-scale \textit{ab-initio} calculations of homobilayers were performed for commensurate twist angle $\theta \sim 4-5^\circ$, and the parameters of an effective continuum model describing the bilayer were fitted to the obtained band structure.
The resulting continuum model, used to describe $\pm K$-pockets $C_3$-symmetric moir\'e semiconducting homobilayers~\cite{wu2019topological,pan2020band,devakul2021magic}, is related to the one obtained in Eq.~\ref{eq_downfoldedmodelwithgradient} by setting $\lambda = \beta = 0$ and treating $(V',W',\psi')$ as free parameters, where primed variable are used to differentiate from our models.

Neglecting the layer-orbit coupling terms in Eq.~\ref{eq_downfoldedmodelwithgradient} to put both models on the same level, we can identify $W' = W (1 - i \beta)$ up to an irrelevant global phase that can be gauged away. 
This justifies treating $V'$ and $W'$ as independent parameters of the theory, but also allows to infer the typical values for $\beta$ from the fitted parameters using $\beta \simeq |W'/\sqrt{V' E_{\rm kin}} -1|$ with $E_{\rm kin} = |\hbar \kappa_0|^2 / 2m^*$. 
These estimates are given in Tab.~\ref{tab_alphaestimates}, showing that $\beta \simeq 0.45$ for both MoTe$_2$ and WSe$_2$ at similar twist angles $\sim 4.5-5^\circ$~\cite{devakul2021magic,reddy2023fractional}.

As a final consistency check, we have fitted the full model Eq.~\ref{eq_downfoldedmodelwithgradient}, \textit{i.e.} including layer-orbit coupling, to the band structure of $5.08^\circ$-twisted WSe$_2$ bilayers~\cite{devakul2021magic}. To avoid over-fitting, we have first fixed $\lambda=1$ and used $\alpha$ and $\beta$ as only free parameters, yielding $\beta = 0.54(9)$ in close agreement with Tab.~\ref{tab_alphaestimates} (see Supplementary Note 7). To test the relevance of the layer-orbit coupling term, we finally fitted the band structure leaving $\lambda$ free but keeping $\alpha$ and $\beta$ fixed to the previously obtained values. This analysis provided the non-negligible best-estimate $\lambda = 0.75(8)$ (see Supplementary Note 7), substantiating the non-negligible role of the layer-orbit coupling highlighted above.

Recall that, in TBG, the dimensionless number playing the role of $\beta$ is the ratio between AA and AB tunneling amplitudes $w_0/w_1 \simeq 0.6-0.8$, which is larger than our estimate $\beta \simeq 0.5$. 
Our estimates therefore suggest that twisted TMDs lie closer to the chiral limit than twisted bilayer, providing a simple argument to understand why they may be better hosts for FCIs~\cite{cai2023signatures,zeng2023integer}.
This difference with TBG is mostly attributable to the stiffer lattice of graphene compared to those of TMDs. 
Indeed, the stronger lattice relaxation effects in twisted TMDs largely influence the $\beta$ ratio, because they increase and reduce the inter-layer distance of the homobilayer respectively when identical atoms overlap ($R_h^h$ stacking) --- decreasing the value of $\beta$ --- and when opposite elements do ($R_M^X/R_X^M$ stacking) --- increasing the value of $t$ --- where the change in hopping amplitude is exponential in inter-layer distance variations~\cite{tong2017topological}.

Because such elastic deformations become increasingly dramatic as the twist angle approaches zero~\cite{carr2018relaxation}, the value of $\beta$ is expected to further decrease in TMD homobilayers at lower twist angles, which can be seen from the much smaller value $\beta \simeq 0.2$ obtained for the fitted parameters of Ref.~\cite{wang2024fractional} for $3.9^\circ$-twisted MoTe$_2$. 
This argument also reveals that smaller elastic constants are desirable to reach the first massive chiral limit and realize its topological flat band physics in the most pristine fashion.
Amongst all semiconducting TMD monolayers with hexagonal lattice, MoTe$_2$ features the lowest Young modulus~\cite{zeng2015electronic,kastuar2022efficient} suggesting its twisted homobilayer form is a slightly better candidate for the emergence of topological correlated phases.

\begin{figure}
\centering
\includegraphics[width=\columnwidth]{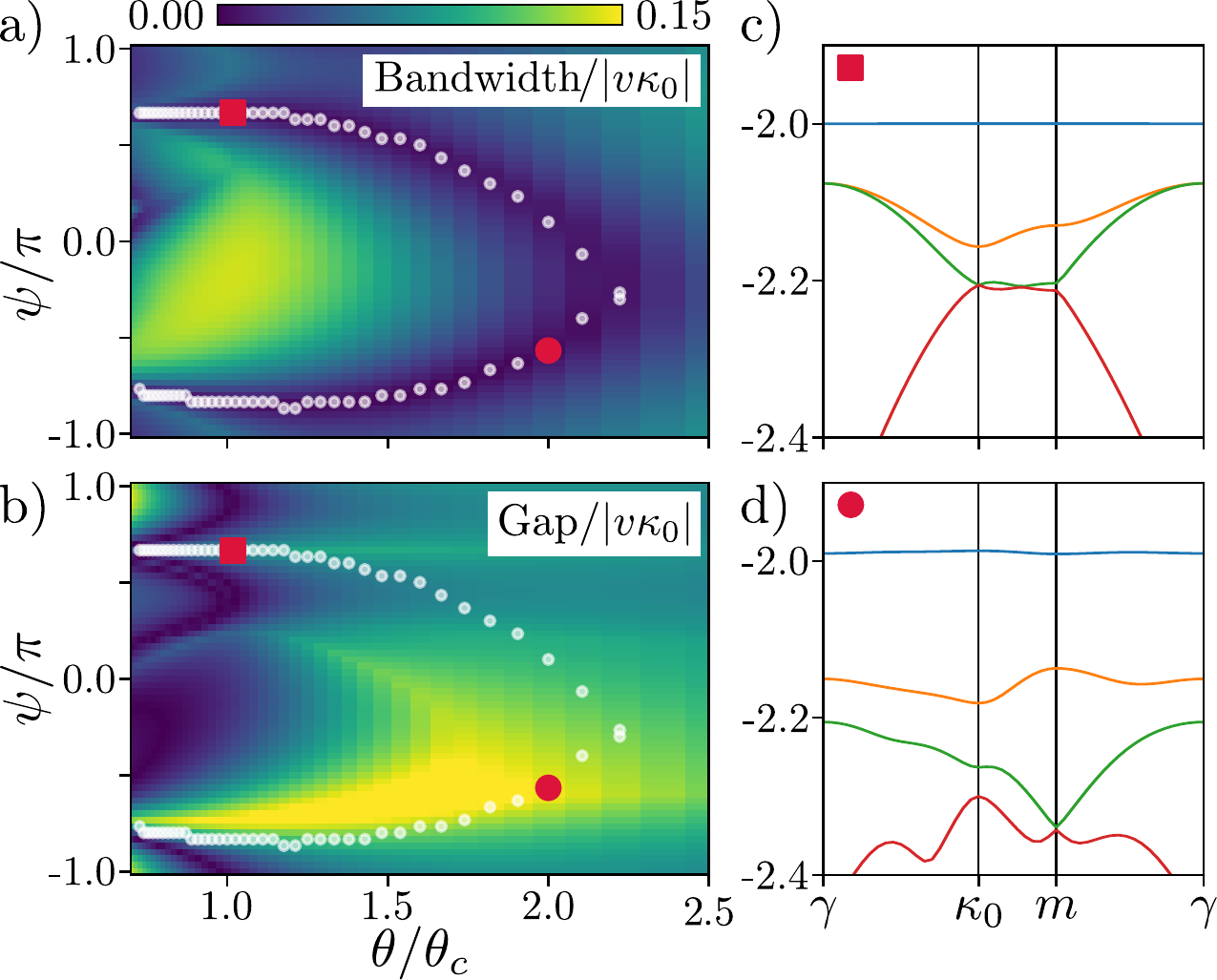}
\caption{Localization of the magic angles for different TMDs. 
a) Bandwidth of the topmost valence band, and (b) direct gap between the two highest valence bands as a function of the twist angle $\theta$ and the angle $\psi$, using $m = 2 |v \kappa_0|$. Minima of the bandwidth are highlighted with white dots. 
We show in (c) and (d) the flat band appearing at the points respectively identified with a square and a circle in (a-b). 
Current literature estimates suggest that (c) and (d) are respectively relevant for WSe$_2$~\cite{devakul2021magic,morales2023pressure} and MoTe$_2$~\cite{wu2019topological,reddy2023fractional}.
All parameters are fixed by Eq.~\ref{eq_downfoldedmodelwithgradient} except $\psi$.
}
\label{fig_ScanPsi}
\end{figure}

Let us finally comment on the fitted values of $\psi'$ obtained in the literature, which consistently provided $\psi' \simeq -\pi/2$ for MoTe$_2$~\cite{wu2019topological,reddy2023fractional} and $\psi'=2\pi/3$ for WSe$_2$~\cite{devakul2021magic,morales2023pressure}. 
To grasp the physical consequences of this difference, we investigate the model Eq.~\ref{eq_downfoldedmodelwithgradient} with all parameters fixed in the massive first chiral limit except $\psi$. 
Although the PH/chiral symmetry only ensures the presence of exact flat bands when $\psi = 2\pi/3$, we observe that our downfolded model still displays an extremely narrow topmost valence band in a wide range of angles $-5\pi/6 \leq \psi \leq 2\pi/3$ (see Fig.~\ref{fig_ScanPsi}a), albeit for different values of the twist angle $\theta$. 
This feature of the model is experimentally desirable for at least two reasons. 
First, moving along the line of near-zero bandwidth allows to reach parameter regimes for which the gap between the first two valence bands is larger than in the chiral limit (see Fig.~\ref{fig_ScanPsi}b).
Second, the parameter $\theta$ at which the bandwidth almost vanishes are always \textit{larger} than $\theta_c$, leading to larger magic twist angle. 
The results of Fig.~\ref{fig_ScanPsi} and the fitted values for $\psi'$ given above provide a simple qualitative insight explaining why MoTe$_2$ ($\psi' = -\pi/2$) realizes correlated topological physics for twist angles twice as large as WSe$_2$ ($\psi'=2\pi/3$). 
Our magic rule applied to WSe$_2$, for which $\psi'$ agrees with Eq.~\ref{eq_downfoldedmodelwithgradient}, provides a magic angle $\theta_c \simeq 1.6^\circ$ in agreement with Ref.~\cite{devakul2021magic}. 
We thus infer $\theta_c \simeq 3.2^\circ$ as magic angle for MoTe$_2$, which is very close to the values $3.4^\circ$ and $3.7^\circ$ where FCIs have been recently observed~\cite{cai2023signatures,zeng2023integer}.

In summary, we have shown that band flattening and exact magic angles in moir\'e semiconductors can be designed akin to TBG, including the existence of chiral limits. Exact magic angles can be described by chiral anomalies in $(2+0)$d dimensions, which we have fully listed for $C_3$ symmetric moir\'e material featuring (gapped or massive) Dirac cones. Only the first chiral limit possess a non-zero Atiyah-Singer index necessary for the emergence of \textit{exact} flat band. Our estimates suggests that twisted TMDs lie closer to that limit than TBG with a ratio $\beta \simeq 0.5$ smaller than $w_0/w_1 \simeq 0.6-0.8$, providing a natural explanation for the emergence of Landau-level like physics in these systems. 
Owing to significant corrugation effects, twisted TMDs are arguably a more natural realization of the chiral limit than twisted bilayer graphene itself.


\begin{table}
\centering
\begin{tabular}{c||c|c|c|c|c||c}
Monolayer & $m^*$ & $a_0 [\SI{}{\angstrom}]$ & twist [$^\circ$] & $V'$ [meV] & $W'$ [meV] & $\beta$ \\ \hline \hline 
WSe$_2$ \cite{devakul2021magic} & 0.38 & 3.317 & 5.08 & 9 & 18 & 0.465 \\
MoTe$_2$ \cite{reddy2023fractional} & 0.62 & 3.472 & 4.4 & 11.2 & 13.3 & 0.453 \\
MoTe$_2$ \cite{wang2024fractional} & 0.62 & 3.472 & 3.89 & 20.8 & 23.8 & 0.187
\end{tabular}
\caption{Estimate of the parameters in Eq.~\ref{eq_downfoldedmodelwithgradient} from large-scale \textit{ab-initio} calculations.  
The parameters $V'$ and $W'$ obtained in Refs.~\cite{devakul2021magic,reddy2023fractional,wang2024fractional} for twisted transition metal dichalcogenides homobilayers provide an estimate for $\beta$, which plays a role analogous to $w_0/w_1$ in TBG~\cite{bistritzer2011moire}. It agrees with the result $\beta \simeq 0.54(9)$ obtained by a direct fit of the model Eq.~\ref{eq_downfoldedmodelwithgradient} on the DFT data of Ref.~\cite{devakul2021magic}, described in the Supplementary Note 7.}
\label{tab_alphaestimates}
\end{table}

\section*{Data availability}

Data sharing is not applicable to this article as no new data were created or analyzed in this study.	

\section*{Acknowledgements}

N.R. is grateful to B.A. Bernevig for previous collaboration on related works and enlightening discussions. V.C. thanks L. Fu for insightful discussions on closely connected topics, and X. Petitcol for his hospitality during the critical phase of this work. 
N.R. acknowledges support from the QuantERA II Programme that has received funding from the European Union’s Horizon 2020 research and innovation programme under Grant Agreement No 101017733. N.R. were also supported by the European Research Council (ERC) under the European Union’s Horizon 2020 research and innovation programme (grant agreement No. 101020833). Research on topological properties of moiré superlattices is supported as part of Programmable Quantum Materials, an Energy Frontier Research Center funded by the U.S. Department of Energy (DOE), Office of Science, Basic Energy Sciences (BES), under award DE-SC0019443. 
The Flatiron Institute is a division of the Simons Foundation.

\section*{Author contributions}

V.C., N.R., and R.Q. conceived of the presented idea. V.C. performed the analytical calculations. All authors discussed the results and contributed to the final manuscript.

\section*{Competing interests}

The authors declare no competing interests.

\section*{References}

\bibliography{magicTMD}

\iftrue

\onecolumngrid
\newpage
\makeatletter 

\begin{center}
\textbf{\large Supplementary material for `` \@title ''}

\vspace{10pt}

Valentin Cr\'epel$^1$, Nicolas Regnault$^{2,3}$ and Raquel Queiroz$^{1,3}$ 

\textit{${}^1$Center for Computational Quantum Physics, Flatiron Institute, New York, New York 10010, USA} 

\textit{${}^2$Laboratoire de Physique de l'Ecole normale sup\'{e}rieure, ENS, Universit\'{e} PSL, CNRS, Sorbonne Universit\'{e}, Universit\'{e} Paris-Diderot, Sorbonne Paris Cit\'{e}, 75005 Paris, France} 

\textit{${}^3$Department of Physics, Columbia University, New York, NY 10027, USA}
\end{center}
\vspace{20pt}

\setcounter{figure}{0}
\setcounter{section}{0}
\setcounter{equation}{0}
\renewcommand{\thesection}{Supplementary Note \@arabic\c@section}
\renewcommand{\theequation}{E\@arabic\c@equation}
\renewcommand{\figurename}{Supplementary Figure }
\makeatother

\twocolumngrid

\appendix

\newpage

\section*{Supplementary Note 1: \\ Unitary transformation and action} \label{app_unitarytransformation}

In this appendix, we show that the action given in Eq.~2 of the main text indeed captures the physics of the Hamiltonian given in Eq.~1 of the main text, that we rewrite here for clarity
\begin{equation} \begin{split}
h (r) & = \sum_{\eta = \pm} \frac{\tau^0 + \eta \tau^3}{2} \sigma^\mu D_\mu^\eta - ( \tau^1 t_\mu \sigma_\mu +  \tau^2 \lambda_\mu \sigma_\mu ) , \\
D_\mu^\pm & = [\pm \delta \mu, i(v \pm \delta v) \partial_1, i (v \pm \delta v) \partial_2 , (m \pm \delta m)] ,
\end{split} \end{equation}
where, to avoid certain minus signs, the indices of Pauli matrices have been lowered as $\sigma_\mu = \eta_{\mu\nu} \sigma^\nu$ using a time-like Minkowski metric $\eta = \diag(+,-,-,-)$. 
\begin{equation}
h' = U h U^\dagger , \quad U = \frac{1}{\sqrt{2}} \begin{bmatrix} - \sigma^0 & \sigma^0 \\ \sigma^3 & \sigma^3 
\end{bmatrix} ,
\end{equation}
which acts on the $\tau^\mu \sigma^\nu$ coefficients as
\begin{equation}
\begin{tabular}{c|c}
\hline
$\mathcal{O}$ & $U \mathcal{O} U^\dagger$ \\
\hline
$\tau^0 \sigma^{1/2}$ & $\tau^3 \sigma^{1/2}$ \\
$\tau^1 \sigma^{1/2}$ & $-\tau^0 \sigma^{1/2}$ \\
$\tau^2 \sigma^{1/2}$ & $(+/-) \tau^1 \sigma^{2/1}$ \\
$\tau^3 \sigma^{1/2}$ & $(-/+) \tau^2 \sigma^{2/1}$ \\
\hline   
\end{tabular} \quad \begin{tabular}{c|c}
\hline
$\mathcal{O}$ & $U \mathcal{O} U^\dagger$ \\
\hline
$\tau^0 \sigma^{0/3}$ & $\tau^0 \sigma^{0/3}$ \\
$\tau^1 \sigma^{0/3}$ & $-\tau^3 \sigma^{0/3}$ \\
$\tau^2 \sigma^{0/3}$ & $-\tau^2 \sigma^{3/0}$ \\
$\tau^3 \sigma^{0/3}$ & $-\tau^1 \sigma^{3/0}$ \\
\hline   
\end{tabular} .
\end{equation}
This leads to 
\begin{align} 
h' &= i v \tau^3 (\sigma^a \partial_a)_{a=1,2} - \tau^0 (\sigma^a t_a)_{a=1,2} + \tau^3 (\sigma^0 t_0 - \sigma^3 t_3) \notag \\ 
& - i (\delta v) \tau^2 (\vec \nabla \times \vec\sigma)_z + \tau^1 (\vec\lambda \times \vec \sigma)_z - \tau^2 (-\lambda_0 \sigma^3 + \lambda_3 \sigma^0) \notag\\ 
& + m \tau^0 \sigma^3 - (\delta m) \tau^1 \sigma^0 - \delta \mu \tau^1 \sigma^3 , 
\end{align}
which, upon introducing the gamma matrices 
\begin{equation}
\gamma^0 = \tau^1 \sigma^0 , \;\, \gamma^{j=1,2,3} = i\tau^2 \sigma^j , \;\, \gamma^5 = i \gamma^0 \gamma^1\gamma^2\gamma^3 = -\tau^3 \sigma^0 ,
\end{equation}
satisfying the Clifford algebra $\{ \gamma^\mu , \gamma^\nu \} = 2 \eta^{\mu\nu}$, simplifies to
\begin{align}
\gamma^0 h' = & - \gamma^a [ i(v -\delta v \gamma^3 \gamma^5) \partial_a + t_a \gamma^5 + i \lambda_a \gamma^3 ] \notag \\ 
& - \gamma^0 ( \delta \mu \gamma^3 \gamma^5 + t_0 \gamma^5 + i\lambda_0 \gamma^3) \notag \\ 
& + t_3 \gamma^3 + i\lambda_3 \gamma^5 + m \gamma^3 \gamma^5 - \delta m  , 
\end{align}
with $a=1,2$. This yields the action
\begin{equation}
\mathcal{S} = \int {\rm d}^3 x \, \psi^\dagger (i v \partial_0 - h' ) \psi ,
\end{equation}
given in Eq.~2 of the main text.

\section*{Supplementary Note 2: \\ Two chiral limits} \label{app_TwoChiralLimits}

In this appendix, we consider all possible chiral limits that exists for the action Eq.~2, and find six of them. However, many become equivalent when the moir\'e pattern has $C_3$ symmetry and the lowest harmonics of the layer hybridization are non-zero, which is currently the most standard experimental setup for moir\'e heterostructures. Note the higher harmonics compatible with the $C_3$ symmetry do not spoil our result, but can change the precise value of the magic angle. In that case, we only find two inequivalent chiral limits for homobilayers, which correspond to the first and second chiral limit of TBG~\cite{bernevig2021twisted,crepel2023chiral,song2021twisted}. Recovering these two known limits roots our method for designing chiral bands in extremely firm grounds.

\subsection{Chiral symmetry candidates}

A chiral symmetry is represented by an operator $\Lambda$ that anti-commutes with all terms bracketed in the action appearing in Eq.~2. Naturally, this requires some of the coefficients to vanish. Here, we first identify all physically relevant candidates and provide conditions on the model for the symmetry to be exact. 

Some of the terms in the original model cannot be set to zero without losing the physical content of the theory. To describe coupled Dirac cone physics, we must have $v\neq 0$ or $\delta v\neq 0$ and at least one of the $t_\mu$ or $\lambda_\mu$ non-zero. Denoting products of gamma matrices as $\Lambda^{i_1, i_2, \cdots i_n} = \gamma^{i_1} \gamma^{i_2} \cdots \gamma^{i_n}$, the operators anticommuting with terms proportional to $v$ in the action and with at least one interlayer coupling are limited to
\begin{equation}
\Lambda \in \{ \Lambda^0 , \Lambda^3 , \Lambda^5 , \Lambda^{035} \} .
\end{equation}
Those anticommuting with terms proportional to $\delta v$ in the action and with at least one interlayer coupling are limited to
\begin{equation}
\Lambda \in \{ \Lambda^0 , \Lambda^{03} , \Lambda^{05} , \Lambda^{035}  \} .
\end{equation}

We now list all terms that must vanish for the identified candidates to be promoted to anticommuting symmetries of the dimensionally reduced action 
\begin{equation} \label{appeq_chiralsymconstraints}
\begin{tabular}{c|c}
Imposed symmetry & Vanishing coefficients \\ \hline 
$\Lambda^0$ & $t_1$, $t_2$, $\lambda_1$, $\lambda_2$, $\delta \mu$, $m$, $\delta m$ \\
$\Lambda^3$ & $t_\mu$, $\delta v$, $\delta \mu$, $\delta m$ \\
$\Lambda^5$ & $\lambda_\mu$, $\delta v$, $\delta \mu$, $\delta m$ \\
$\Lambda^{03}$ & $t_1$, $t_2$, $\lambda_0$, $\lambda_3$, $v$, $\delta \mu$, $\delta m$ \\
$\Lambda^{05}$ & $t_0$, $t_3$, $\lambda_1$, $\lambda_2$, $v$, $\delta \mu$, $\delta m$ \\
$\Lambda^{035}$ & $t_0$, $t_3$, $\lambda_0$, $\lambda_3$, $\delta \mu$, $m$, $\delta m$ 
\end{tabular} .
\end{equation}
Note that these terms should only vanish in the effective $(2+0)$d action where the chiral anomaly occurs. This allows us, for instance, to compensate for a mass term using $\bar \psi v \gamma^0 \partial_0 \psi = E \psi^\dagger \mathcal{O} \psi$ with $E = m$ and $\mathcal{O} = \gamma^0 \gamma^3 \gamma^5$.

\subsection{$C_3$ and non-zero lowest harmonics}

For concreteness, we now focus on the most common class of moir\'e materials studied in the main text: $C_3$ symmetric moir\'e materials possessing massless or massive Dirac cones at the corner of their Brillouin zone.

Approximating the long-wavelength moir\'e potentials using the three lowest harmonics allowed by the moir\'e pattern leads to
\begin{equation} \label{appeq_lowestharmonicsC3}
T(r) = \sum_{n=0,1,2} T_n e^{i(\kappa_n \cdot r)} ,
\end{equation}
where the coefficients in $T_n$ have a well defined $C_3$ eigenvalues, $T_n^{ij} = \omega^{\nu_{ij} n} T_0^{ij}$ where $\nu_{ij} \in \{-1,0,1\}$ a $C_3$ eigenvalue. In fact, for the full spectrum of the original model Eq.~1 to be $C_3$ invariant, we must have $\nu_{00} = \nu_{11} = 0$ and $\nu_{01} = -\nu_{10} =1$~\cite{bistritzer2011moire,bernevig2021twisted,song2021twisted,tarnopolsky2019origin}.

The form of Eq.~\ref{appeq_lowestharmonicsC3} requires any non-zero coefficient $T_{ij}$ to have both non-zero real and imaginary components -- a peculiarity of the two inequivalent $K$ points of the Brillouin zone. 
This enforces more stringent constraints on the models when combined with Eq.~\ref{appeq_chiralsymconstraints}
\begin{equation} \label{appeq_chiralsymconstraints2}
\begin{tabular}{c|c}
Imposed symmetry & Vanishing coefficients \\ \hline 
$\Lambda^0$ & $t_{1,2}$, $\lambda_{1,2}$, $\mu$, $\delta \mu$, $m$, $\delta m$ \\
$\Lambda^3$ & $t_\mu$, $\lambda_{0,3}$, $\delta v$, $\delta \mu$, $\delta m$ \\
$\Lambda^5$ & $\lambda_\mu$, $t_{0,3}$, $\delta v$, $\delta \mu$, $\delta m$ \\
$\Lambda^{03}$ & $t_\mu$, $\lambda_{0,3}$, $v$, $\delta \mu$, $\delta m$ \\
$\Lambda^{05}$ & $\lambda_\mu$, $t_{0,3}$, $v$, $\delta \mu$, $\delta m$ \\
$\Lambda^{035}$ & $t_{0,3}$, $\lambda_{0,3}$, $\mu$, $\delta \mu$, $m$, $\delta m$ 
\end{tabular} .
\end{equation}

The model obtained for $\Lambda^3$ is related by a unitary gauge transformation $U = [(\tau^0+\tau^3) \sigma^0 + (\tau^0-\tau^3) \sigma^3]/2$ to the one obtained after imposing the chiral symmetry $\Lambda^5$. Similarly, the models for $\Lambda^{03}$ and $\Lambda^{05}$ are related by the same transformation. 
We further notice that both the $\Lambda^5$ and $\Lambda^{05}$ models feature a hermitian matrix $T(r) = t_1 \sigma^1 + t_2 \sigma^2$, which is incompatible with the form of Eq.~\ref{appeq_lowestharmonicsC3} unless $T=0$, \textit{i.e.} unless the layer are decoupled and there is no heterostructure. We can thus discard both of these limits and their equivalents from now on. As a result, there are only two distinct chiral limits for moir\'e Dirac materials, described by the operators $\Lambda^{035}$ and $\Lambda^{0}$.

\subsection{Homobilayers}

Homobilayers enjoy an additional symmetry: the inversion of the stacking registry and layer index together must leave the system unchanged. This ensures that $\delta v = 0$ and $T^\dagger (-r) = T(r)$, which leads to the additional requirement $|T_0^{ij}|=|T_0^{ji}|$. 
We finally look at the consequences of this last relation on the two chiral limits identified above, which both require $\delta \mu = m = \delta m =0$ in the dimensionally reduced $(2+0)$d action (Eq.~\ref{appeq_chiralsymconstraints2}).

For $\Lambda^{035}$, we find that the inter-layer hybridization is of the form 
\begin{equation} \label{appeq_tunnelingfirstchirallimit}
T^{\rm st}(r) = t \sum_{n=0}^2 e^{i \kappa_n \cdot r} \begin{bmatrix} 0 & \omega^n \\ \omega^{-n} & 0 \end{bmatrix} , \quad \omega = e^{2i\pi/3} ,
\end{equation}
which corresponds to the first chiral limit of TBG~\cite{bernevig2021twisted,lian2021twisted}. 
For $\Lambda^0$, we get the diagonal form 
\begin{equation} \label{appeq_tunnelingsecondchirallimit}
T^{\rm nd}(r) = t \sum_{n=0}^2 e^{i \kappa_n \cdot r} \begin{bmatrix} 1 + \epsilon & 0 \\ 0 & 1-\epsilon \end{bmatrix} .
\end{equation}
These are the forms provided in Eqs.~3 and~4 of the main text. Note that the value of $\epsilon$ is irrelevant to the physics described in the main text, since we focus on the valence or conduction degrees of freedom only but never take both into account at the same time. We thus set $\epsilon=0$ from now on.

\section*{Supplementary Note 3: \\ No flat bands in the second chiral limit} \label{app_IndexSecondChiralLimit}

In this appendix, we compute the Atiyah-Singer index in the second chiral limit and show that it cannot be a non-zero integer. The consistency condition for flat bands in the second chiral limit is never satisfied, showing that no exact flat bands exist in the second chiral limit.

The calculation of the Atiyah-Singer index proceeds along the lines of Ref.~\cite{parhizkar2023generic} --- to which we refer for details on the Fujikawa method~\cite{fujikawa1979path}. To treat both chiral limits within the same framework, we apply the unitary transformation $U'= [(\tau^0+\tau^3) \sigma^0 + (\tau^0-\tau^3) \sigma^1]/2$ to the model Eq.~1 obtained in the second chiral limit, which brings the tunneling to off-diagonal form $T \to \sigma^1 T$, with the identification $(t_{0} , t_3 , \lambda_0 , \lambda_3) \to ( - t_1, - \lambda_2 , - \lambda_1, t_2)$ between the coefficients. As a result, the fields carried by $\gamma^5$ and $\gamma^3$ in these two chiral limits can be represented by 
\begin{equation} \label{appeq_chiralsymconstraints3}
\begin{tabular}{c|c|c}
 & first chiral limit & second chiral limit \\ \hline 
$t_1$ & $\frac{t}{2} [2c_0 - c_1 -c_2]$ & $-t \sum_n c_n$ \\
$t_2$ & $\frac{\sqrt{3}t}{2} [c_2-c_1]$ & $0$ \\
$\lambda_1$ & $\frac{t}{2} [2s_0 - s_1 -s_2]$ & $-t \sum_n s_n$ \\
$\lambda_2$ & $\frac{\sqrt{3}t}{2} [s_2-s_1]$ & $0$ \\
\end{tabular} ,
\end{equation}
with $c_n = \cos \kappa_n \cdot r$ and $s_n = \sin \kappa_n \cdot r$. The corresponding field strength $F = (\partial_2 t_1 - \partial_1 t_2)/v$~\cite{parhizkar2023generic} takes the form 
\begin{align} \label{appeq_fictitiousfield}
F^{\rm st} & = - \frac{t|\kappa_0|}{v} \sum_{n=0,1,2} \sin(\kappa_n \cdot r) , \\
F^{\rm nd} & = \frac{t|\kappa_0|}{v} \left[ \sin(\kappa_0 \cdot r) -\frac{1}{2}\sin(\kappa_1 \cdot r) -\frac{1}{2} \sin(\kappa_2 \cdot r) \right]  \notag .
\end{align}

\begin{figure}
\centering
\includegraphics[width=0.75\columnwidth]{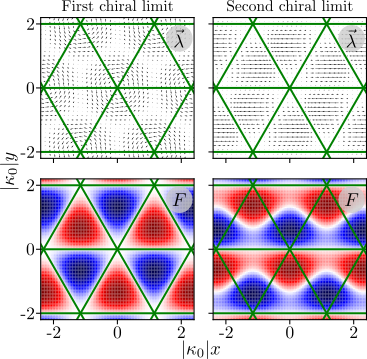}
\caption{Vector field $\vec\lambda$ (top) and fictitious field strength $F$ (bottom) in the first (left) and second (right) chiral limits. The elementary patches on which the Atiyah-Singer index is computed (green) are defined by the condition that $\vec\lambda$ either vanishes on or is orthogonal to their boundaries.}
\label{fig_ASindex}
\end{figure}

The Atiyah-Singer index can be obtained as the flux of this fictitious magnetic field $F$ through an elementary patch $\triangle$ in real-space, $n_{\rm AS} = \int_{\triangle} {\rm d} x^2 F$, where the contour of $\triangle$ are chosen such that the vector $\vec\lambda = \lambda_{a=1,2}$ either is perpendicular to (first chiral limit) or vanishes on (second chiral limit) the boundary of the patch~\cite{parhizkar2023generic}, as shown in Supplementary Figure ~\ref{fig_ASindex}.

The $2\pi/3$ rotation invariance of the problem around the center of each patch ensures that all the sines (or cosines) in $F$ give the same integral contribution to $n_{\rm AS}$. 
It is then clear from Eq.~\ref{appeq_fictitiousfield} that the first chiral limit has a non-zero Atiyah-Singer while this index \textit{always} vanishes in the second chiral limit. 
Performing the integral on a patch where $F^{\rm st}>0$, we find 
\begin{equation}
n_{\rm AS}^{\rm st} = \frac{\sqrt{3}t}{|v \kappa_0|} ,  \quad n_{\rm AS}^{\rm nd} = 0 .
\end{equation}
The consistency condition for the emergence of flat bands is never met in the second chiral limit, while it reads $\alpha =  t/|v \kappa_0| = n/\sqrt{3}$, with $n$ a positive integer, in the first chiral limit~\cite{tarnopolsky2019origin,parhizkar2023generic}. 
This sets the value $\alpha_c = 1/\sqrt{3}$ used in the main text.

\newpage

\section*{Supplementary Note 4: \\ Landau level-like flat bands} \label{app_bandsareideal}

\begin{figure}
\centering
\includegraphics[width=0.85\columnwidth]{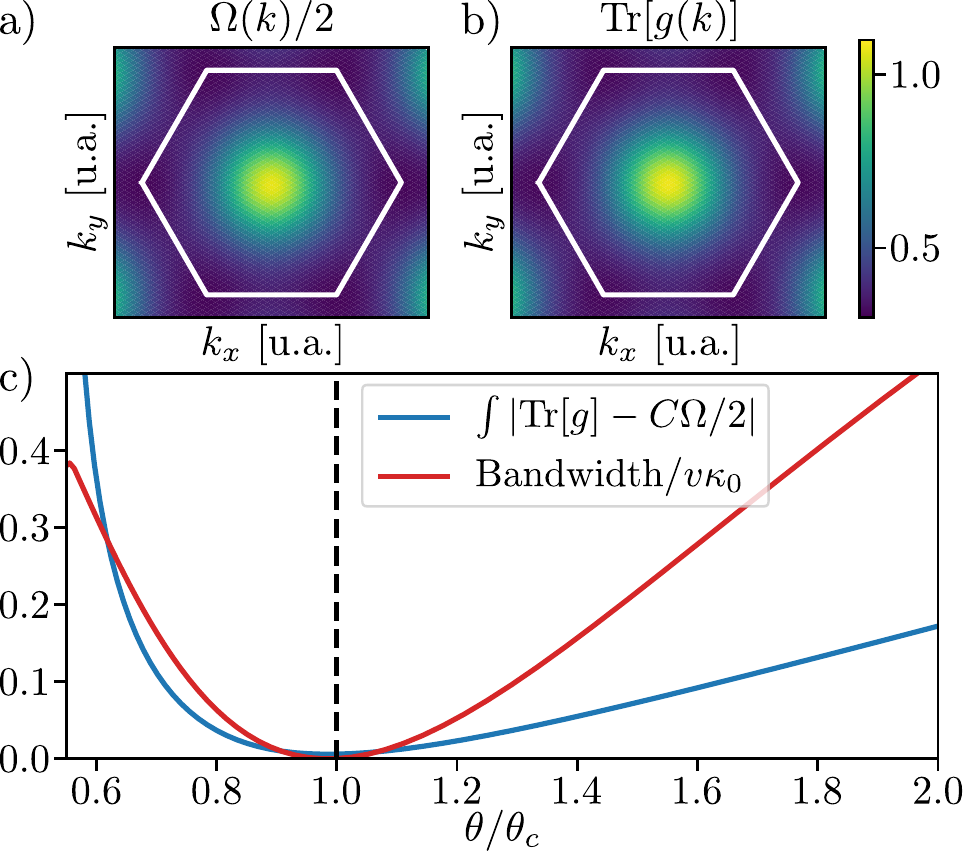}
\caption{The Berry curvature (a) and the trace of the quantum metric (b) of the downfolded model Eq.~5 perfectly match at the magic angle $\theta =\theta_c$. c) Away form the magic angle, both the bandwidth and deviation of the trace conditions become nonzero, as shown here in the chiral limit $\beta =0$.}
\label{fig_idealness}
\end{figure}

The flat bands of TBG in the first chiral limit at the first magic angle are known to mimic the physics of a Landau level in presence of a non-uniform magnetic field~\cite{wang2021chiral}, in the sense that one can find a Landau problem that exactly reproduces all form factors of the flat bands~\cite{estienne2023ideal} providing a fertile playground for fractional quantum Hall-like phases to arise~\cite{tarnopolsky2019origin,wang2021chiral,crepel2018matrix}. This property of the band, dubbed ``$q$-idealness", is probed by the trace condition~\cite{wang2021exact}
\begin{equation}
{\rm Tr} [ g(k) ] = C \Omega (k) /2 , 
\end{equation}
where we have introduced the Berry curvature $\Omega(k)$, the Chern number $C = (2\pi)^{-1} \int_{BZ} \Omega(k) {\rm d}^2 k$ and the quantum metric $g_{ab}(k)$ through $\braket{D_k^a u_k}{D_k^b u_k} = g^{ab}(k) + i \varepsilon^{ab} \Omega(k)/2$ with $\ket{u_k}$ the periodic part of the Bloch wavefunctions and $\ket{D_k^a u_k} = \ket{\partial_{k_a} u_k} - \braOket{u_k}{\partial_{k_a}}{u_k} \ket{u_k}$. We have numerically checked that this condition was also satisfied at $\alpha = \alpha^*$ in the downfolded model Eq.~5 of the main text, as shown in Supplementary Figure~\ref{fig_idealness}.

\section*{Supplementary Note 5: \\ Mass terms and downfolding} \label{app_downfolding}

In this appendix, we consider the massive first chiral limit and downfold its Hamiltonian onto the valence band of the semiconducting layers, assuming the mass to largely dominate all other energy scales of the model. We recover the standard continuum model for moir\'e TMD bilayer~\cite{wu2019topological,crepel2023topological} with the addition of another term taking the form of a layer-orbit coupling.

To that aim, we first rewrite the Hamiltonian of Eq.~1 of the main text in the first chiral limit using a basis that more naturally separates high and low degrees of freedoms
\begin{equation}
h^{(2)} = U' h (U')^\dagger , \quad U' = \begin{bmatrix} 1&0&0&0\\0&0&1&0\\0&1&0&0\\0&0&0&1 \end{bmatrix} , 
\end{equation}
leading to
\begin{align}
h^{(2)} (r) & = \begin{bmatrix} m \tau^0 &0 \\ 0 & - m\tau^0 \end{bmatrix} +  \begin{bmatrix} 0&D_+ \\ D_-&0 \end{bmatrix} , \\ D_+ = D_-^\dagger &= \begin{bmatrix} iv (\partial_1 -i \partial_2) & T_{10}^* \\ T_{01} & iv (\partial_1 -i \partial_2)  \end{bmatrix} ,   \notag
\end{align}
where we recall that (see Eq.~\ref{appeq_tunnelingfirstchirallimit})
\begin{equation}
T(r) = \begin{bmatrix} 0 & T_{01} \\ T_{10} & 0 \end{bmatrix} = t \sum_{n=0}^2 e^{i (\kappa_n \cdot r)} \begin{bmatrix} 0 & \omega^{n} \\ \omega^{-n} & 0 \end{bmatrix} .
\end{equation}
Standard perturbation theory yields, up to second order corrections in the large mass term $m$, the following approximation of $h^{(2)}$
\begin{equation}
h^{(3)} (r) \simeq \begin{bmatrix} m \tau^0 + \frac{1}{2m} D_+ D_- & 0 \\ 0 & - m \tau^0 - \frac{1}{2m} D_- D_+ \end{bmatrix} ,
\end{equation}
in which positive and negative energy states are effectively decoupled to second order in the large mass $m$. 
Note that the off diagonal block structure of $h^{(2)}$ offers an interesting relation 
\begin{equation}
h^{(3)} = \frac{1}{2} \begin{bmatrix} \tau^0 &0 \\ 0 & -\tau^0 \end{bmatrix} \left( m + \frac{[h^{(2)}]^2}{m}  \right) , 
\end{equation}
such that the effective Hamiltonian $h^{(3)}$ can be simply understood as the square of the original Hamiltonian.

We now focus on negative energies, as in the main text, and give a more explicit form of the downfolded Hamiltonian 
\begin{equation}
\tilde h  = -m \tau^0 - \frac{1}{2m} D_- D_+ .
\end{equation}
Direct evaluation of all terms appearing in $D_- D_+$ using the functional form of $T(r)$ gives
\begin{equation}
\tilde h  = \left[ - m + \frac{v^2 \nabla^2}{2m} \right] \tau^0 - \begin{bmatrix} a_+ & b^* \\ b & a_- \end{bmatrix}  ,  
\end{equation}
where
\begin{equation} 
a_+ = \frac{|T_{01}|^2}{2m} , \;\, a_- = \frac{|T_{10}|^2}{2m} , \;\, b  = \frac{iv}{m} \left[ T_{10} \partial_z + \bar\partial_z ( T_{01} \cdot ) \right] \!.
\end{equation}
The diagonal coefficients take the simple form 
\begin{equation}
a_\tau = V \left[ 3 + 2 \sum_{n=0,1,2} \cos(g_n \cdot r + \tau \psi) \right] , \quad V = \frac{|t|^2}{2m} ,
\end{equation}
with $g_n = \kappa_n - \kappa_{n-1}$ the primitive moir\'e reciprocal lattice vector(see Fig.~1 of the main text); and the off-diagonal ones read 
\begin{widetext} \begin{equation}
b = i W \sum_{n=0,1,2} e^{i (\kappa_n \cdot r)} \left[ 1 +  \frac{(\omega^n + \omega^{-n})\partial_1 + i (\omega^n - \omega^{-n})\partial_2]}{|\kappa_0|} \right] = i W \sum_{n=0,1,2} e^{i (\kappa_n \cdot r)} \left[ 1 +  \frac{2 (\kappa_n \times \nabla)}{|\kappa_0|^2} \right] , \quad W = \frac{t |v \kappa_0|}{2m} ,
\end{equation} \end{widetext}
where the global $i$ factor comes from our convention that $\kappa_0$ points along $(-\hat y)$, and the second form of the gradient term (used in the main text) is manifestly $C_3$ symmetric and comes from 
\begin{equation}
\kappa_n = \frac{|\kappa_0|}{2} [i (\omega^n - \omega^{-n}), -(\omega^n + \omega^{-n})] .
\end{equation}

Note that we can very easily consider deviations away from the first chiral limit with a interlayer hybridization matrix $T = T^{\rm st} + \frac{|v\kappa_0|}{2m} \beta T^{\rm nd}$ because $T^{\rm nd}$ is diagonal and thus does not mix the conduction and valence bands. 
As a result, it leaves the second order perturbation unchanged to second order, and can be simply add to the previously obtained effective Hamiltonian. 
This results in 
\begin{equation}
\tilde h = \left[ - m + \frac{v^2 \nabla^2}{2m} \right] \tau^0 - \begin{bmatrix} a_+ & b^*  \\ b & a_- \end{bmatrix}  + W \beta (1-\epsilon) \sum_n e^{i \kappa_n \cdot r} \tau^1 .
\end{equation}
Because we trace over the conduction degrees of freedom, the only effect of $\epsilon$ is to slightly renormalize the inter-layer valence-valence tunneling coefficient, which we can also account for through the redefinition $\beta (1-\epsilon) \to \beta$ and $\epsilon \to 0$. 
This shortens the notations and yields the Eq.~5 of the main text. 

\newpage

\section*{Supplementary Note 6: \\ Rule for magic} \label{app_ruleofthumb}

\begin{figure}
\includegraphics[width=\columnwidth]{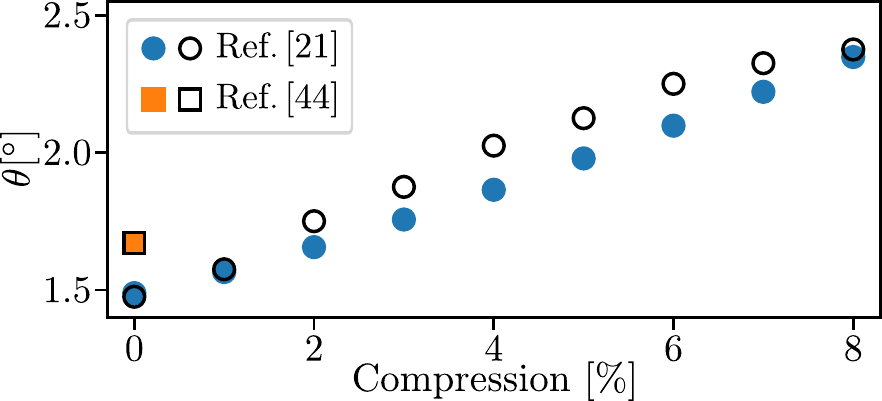}
\caption{
Comparison between the magic angle predicted by Eq.~6 and the data obtained for twisted WSe$_2$ homobilayers in absence of strain in Ref.~\cite{devakul2021magic} and in presence of a uniform uni-axial compression $x$ of the bilayer in Ref.~\cite{morales2023pressure}. 
Empty symbols shows the exact the magic angle predicted by Eq.~6 for the parameters $(V',W')$ given in Eq.~\ref{eq_parametersfrompreviousrefs}. 
Filled symbols are the twist angles at which the variance of Berry curvature is minimal in the continuum model of Refs.~\cite{devakul2021magic,morales2023pressure} for the same values of $(V',W')$. 
}
\label{fig_matchWse2}
\end{figure}

As discussed in the main text, fit to large-scale \textit{ab-initio} band structures of commensurate WSe$_2$ homobilayers~\cite{devakul2021magic,morales2023pressure} provided values of $\psi'$ near $2\pi/3$, agreeing with Eq.~5 of the main text. 
The other two parameters of their continuum model were determined both in absence of strain (Ref.~\cite{devakul2021magic})
\begin{subequations} \label{eq_parametersfrompreviousrefs} \begin{equation}
V' = \SI{9}{\milli\electronvolt} , \quad W' = \SI{18}{\milli\electronvolt} , 
\end{equation}
and in the presence of a uniform uni-axial compression $x$ of the bilayer (Ref.~\cite{morales2023pressure})
\begin{equation} \begin{split}
V' &= (6.359 + 1.3982 x+0.0456 x^2) {\rm meV} , \\ 
W' &= (8.897 + 1.3860 x+0.0811 x^2) {\rm meV} , 
\end{split} \end{equation} \end{subequations}
where $x$ is measured in percent. 
Using these two parameters $(V',W')$, together with the effective mass and the lattice constant provided in Tab.~I of the main text, the magic rule of Eq.~6 in the main text (obtained in the first massive chiral limit with $\beta =0$) allows us to predict a magic angle for the twisted WSe$_2$ homobilayers considered in Refs.~\cite{devakul2021magic,morales2023pressure}. 
Because there is no exact magic angle in the continuum models used in these references --- it lacks the crucial layer-coupling term $\lambda$ of Eq.~5 --- we choose one possible probe to compare our prediction for the magic angle: the angle at which the system exhibits the lowest Berry curvature variance. 
Our prediction for the magic angle and the angle minimizing Berry curvature fluctuations in Refs.~\cite{devakul2021magic,morales2023pressure} are compared in Supplementary Figure~\ref{fig_matchWse2}, and shows very good agreement. 
Beyond our analytical derivation, this provides another justification of our magic rule Eq.~6 in moir\'e semiconductors.

\newpage

\section*{Supplementary Note 7: \\ Fit to large twist angle data} \label{app_fittolargetwists}

In this appendix, we fit the only two parameters $\alpha$ and $\beta$ of the downfolded Dirac model Eq.~5 on the band structure obtained in Ref.~\cite{devakul2021magic} for twisted WSe$_2$ at $5.08^\circ$~\cite{devakul2021magic} (black symbols in Supplementary Figure~\ref{fig_fitDFT}). We find $\alpha = 0.051$ and $\beta = 0.54$ with a relatively small ($3\%$) statistical uncertainty on $\alpha$ and a much large one ($17\%$) on $\beta$, which is quoted in the main text. The band structure of our continuum model Eq.~5, with the fitted values of $\alpha$ and $\beta$ given above, is shown in Supplementary Figure~\ref{fig_fitDFT}. We also show the band structure of the continuum model used in Ref.~\cite{devakul2021magic} for comparison. At this angle, both models equally well capture the topmost valence band of the \textit{ab-initio} band structure. 

\begin{figure}
\includegraphics[width=\columnwidth]{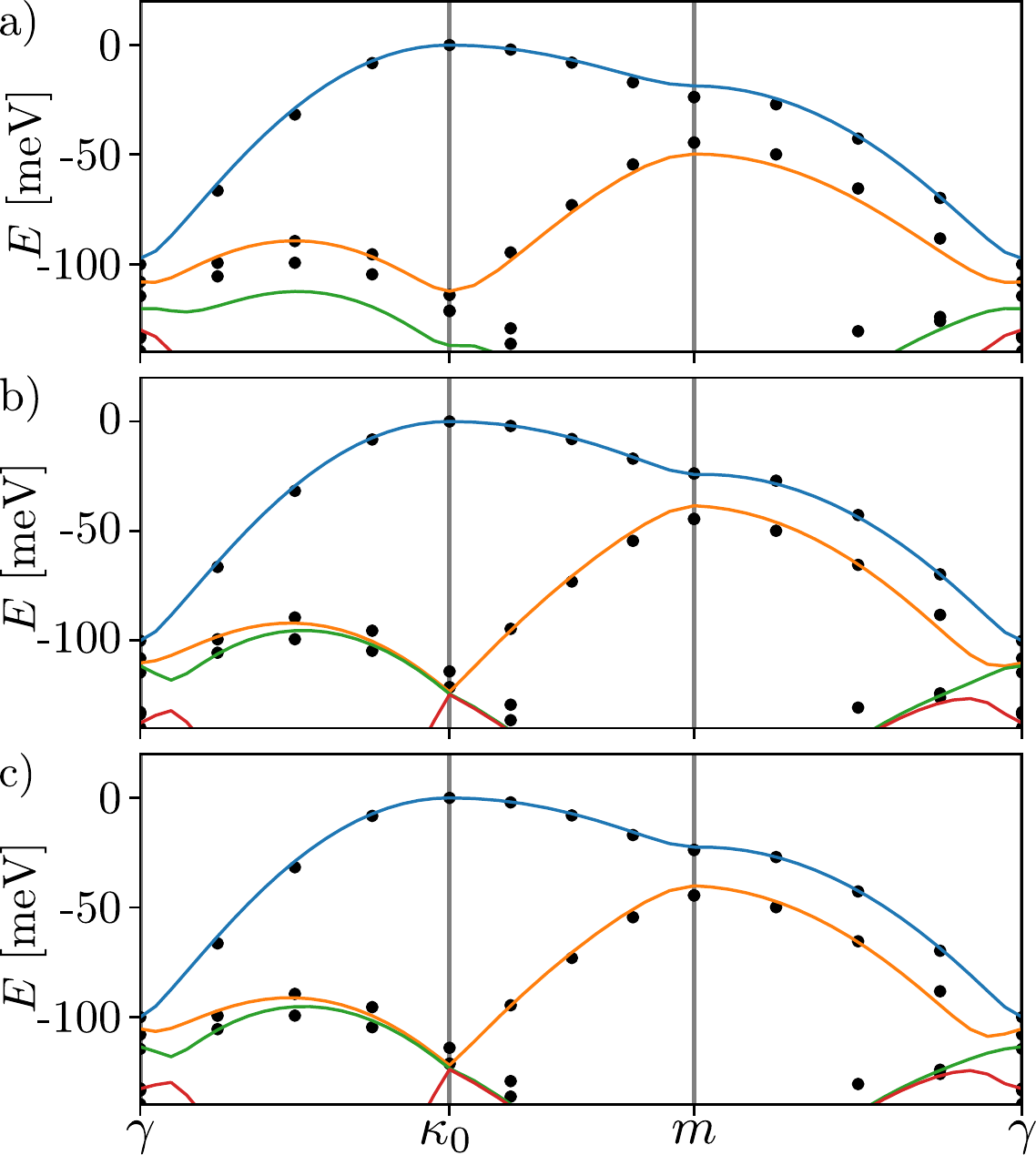}
\caption{In (a), (b) and (c), the black dots represent the band structure  of a $5.08^\circ$-twisted WSe$_2$ homobilayer obtained by large scale \textit{ab-initio} calculations in Ref.~\cite{devakul2021magic}.
Colored lines are the result of (a) the standard continuum model for twisted transition metal dichalcogenides found in Ref.~\cite{devakul2021magic}, (b) the downfolded model Eq.~5 with $\lambda=1$ and $(\alpha,\beta)$ fitted to the \textit{ab-initio} data, and (c) the downfolded model Eq.~5 with $\beta = 0.54$ obtained in (b) and consistent with the estimates of Tab.~I, and $(\alpha,\lambda)$ fitted  to the \textit{ab-initio} data.
}
\label{fig_fitDFT}
\end{figure}

To gauge the importance of the layer-orbit coupling term uncovered in the main text, we finally perform a fit of the \textit{ab-initio} band structure leaving $\alpha$ and $\lambda$ free and fixing $\beta = 0.54$ to the value found above (fitting all parameters at the same time was observed to yield highly correlated results and large statistical uncertainties). This yields a similar value of $\alpha = 0.064$, and a non-negligible $\lambda = 0.70$ with respective statistical error $8\%$ and $11\%$. The corresponding band structure is shown in Fig.~\ref{fig_fitDFT}c.

\fi
\end{document}